\documentclass[aps,pra,floatfix,twocolumn,superscriptaddress]{revtex4-1}
\usepackage{graphicx}
\usepackage{float}
\usepackage{subfigure}
\usepackage[english]{babel}
\usepackage{amsmath}
\usepackage{amssymb}
\usepackage{tensor}

\newcommand{\vk}{\mathbf{k}}

\newcommand{\be}{\begin{eqnarray}}
\newcommand{\ee}{\end{eqnarray}}
\newcommand{\p}{\partial}

\def\ket#1{|#1\rangle}

\def\ep#1{\langle #1 \rangle}

\begin{document}

\title{Local Optical Conductivity of Bilayer Graphene with Kink Potential}

\author{ Zhen-Bing Dai}
\affiliation{College of Physics, Sichuan University, Chengdu, Sichuan 610064, China}
\affiliation{Department of Physics, Sichuan Normal University, Chengdu, Sichuan 610066, China}
\author{Zhiqiang Li}
\affiliation{College of Physics, Sichuan University, Chengdu, Sichuan 610064, China}
\author{Yan He}
\thanks{Electronic address: heyan\_ctp@scu.edu.cn}
\affiliation{College of Physics, Sichuan University, Chengdu, Sichuan 610064, China}

\begin{abstract}
We study the optical response of bilayer graphene with a kink potential composed of a domain wall separating two AB regions with opposite interlayer bias. The band structure and the local optical conductivity in real space are investigated in details based on a continuum model. We find that the one-dimensional chiral states localized at the domain wall contribute significantly to the local optical conductivity, which shows a clear distinction in different regions. The effects of domain wall states on optical conductivity can be detected by spatially and frequency resolved spectroscopic features. From the spectrum at various Fermi energies, important features in the band structure such as the energy separation between two chiral states can be directly measured. When the domain wall region is broad, the spatial distribution of local optical conductivity can provide important information on the bound states as well as the topological domain wall states.
\end{abstract}

\maketitle
\section{Introduction}

In bilayer graphene(BLG), the chiral counterpropagating boundary modes occur at the interface between the AB and BA stacking configurations~\cite{MacDonald11,Vaezi13,JuL15,Walet20}, or at the line junction of two oppositely gated regions~\cite{Martin08,Zarenia11,LiJ16,Chen20}, because the Berry curvatures in these two regions near the domain wall(DW) have opposite signs~\cite{Yao09,ZhangF11,ZhangF13,Song17}. These edge states are topologically protected because the inter-valley scattering is almost absent~\cite{Qiao11}. A triangular network of topologically protected layer stacking DW states has been found in twisted BLG with a tiny twist angle of $\theta=0.06^\circ$ ~\cite{Sunku18}, due to the atomic scale lattice reconstruction~\cite{Gargiulo17}. The area of the DW regions and AB and BA domains become comparable as the twist angle increases to a magic angle($\theta_c=1.05^\circ$), in which superconductivity and Mott insulator-like states~\cite{CaoY1801,CaoY1802} have been observed~\cite{Kim17,MacDonald11}.

Compared to the DW between different layer stackings, DW due to a kink potential in BLG could offer a robust and flexible platform to realize topological and valleytronic operations~\cite{Qiao11,Qiao14, Jung11,Abdullah17}, such as long lived 1D DW plasmons~\cite{Song17} , waveguides~\cite{ZhangFM09}, quantum valley Hall states~\cite{Yin16} and valley filters~\cite{Rycerz07}. In addition, net valley currents can be induced by the counter-propagating conducting channels of the gated BLG~\cite{Gorbachev14,Shimazaki15,Lee17}, which can be detected directly by the measurements of the valley Hall effect. Moreover, the tunability of the dual-split-gating configuration has been conveniently utilized in real devices. Each gate pair can operate independently and induce DW states within the split-gate region, which provides many opportunities for potential device applications.

Optical methods are widely used to quantitatively study the electronic band structure and physical properties of materials. It is reported ~\cite{jiang17,JiangL16,Fei13,Ju17, Yin16} that the surface plasmon reflection at DW solitons can be studied in detail by the near-field infrared nanoscopy for the reversal stacking configurations in BLG. The DW plasmon also displays a rich phenomenology including tunable subwavelength confinement of light, higher propagation length, plasmon oscillation over a wide range of frequencies as well as supporting the propagation of transverse electric modes~\cite{Farzaneh17,Mikhailov07,Song17}. Meanwhile, optical conductivity measurements can provide a feasible way to probe excitation spectrum of BLG between the gapless states and the continuum bands~\cite{Basov11}. Up to now, most works focus on the electronic structure and the frequency dependence of the optical conductivity of homogeneous graphene~\cite{Mccann13,Moon13,Koshino13-2,Farzaneh17,Min11}. The local optical conductivity of BLG is strongly sensitive to the atomic-scale stacking order~\cite{jiang17} and the band structure which associated with electric-field-induced energy asymmetry~\cite{Farzaneh17}, leading to the inhomogeneous distributions across DW. To the best of our knowledge, a systematic theoretical study of the optical properties of opposite gated BLG is still lacking. In this paper, we will provide a comprehensive way to investigate the band structure and the local optical conductivity of opposite gated BLG.

We start with the four-band continuum model of BLG, and discuss the calculation method of electronic structure, topological properties and the related local optical conductivity for the opposite gated BLG. In the numerical calculation, we provide a method to avoiding the fermion doubling problem that is encountered in the lattice Hamiltonian. We analyze the band structure of the gated BLG and the wave functions of a few selected gapless modes. The origin of the gapless modes can be distinguished from the profiles of the wave function. In the end, we present the spatial and frequency distributions of the real part of the local optical conductivity across DW with different chemical potentials.

\section{Theoretical background}

In order to derive and discuss the distributions of the optical conductivity in real space of BLG under the asymmetry electronic potential, we will examine the band structure of gated BLG. To this end, we begin with a continuum model~\cite{Martin08,Mccann13} of BLG with a bias potential applied to these two layers.

The Hamiltonian is given by
\be
&&H_{\xi}=-\frac{V}{2} \tau_3\otimes \sigma_0+v_F k_x \tau_0\otimes\sigma_1+\xi v_F k_y \tau_0\otimes\sigma_2\nonumber\\
&&\quad+\frac{t_{\bot}}{2}(\tau_1\otimes\sigma_1+\tau_2\otimes\sigma_2),
\label{Hk}
\ee
where $(\tau_1,\tau_2,\tau_3)$ and $(\sigma_1,\sigma_2,\sigma_3)$ are Pauli matrices acting on the layer space and sublattice space, respectively. In addition, $\tau_0$ and $\sigma_0$ are the corresponding identity matrices, the band velocity of the Dirac cone $v_F=\frac{\sqrt{3}a\gamma_0}{2\hbar}$ is expressed in terms of the in-plane nearest neighbor interaction $\gamma_0=2.7$ eV~\cite{Moon13,Trambly10} and the lattice constant of monolayer graphene $a\approx0.246$ nm, and $t_{\bot}=0.39$ eV denotes the interlayer coupling term between adjacent layers. The corresponding wave-function is arranged as $\psi=(\psi_{A1},\psi_{B1},\psi_{A2},\psi_{B2})^T$ where $A,B$ label two sublattices and $1,2$ label two layers. Note that the low energy Hamiltonian in Eq.(\ref{Hk}) is for $K \xi$ valley with $\xi=\pm1$. The advantage of the continuum model is that we can focus on a single Dirac cone at one valley without mixing effects of two inequivalent valleys. For simplicity, we set $\hbar=1$ hereafter.

We first consider the topological properties of gated BLG case with constant bias voltage. In this case, one can directly diagonalize the Hamiltonian in momentum space to find out four eigenvectors $\psi_m(\vk)$, where $m=1,\cdots,4$ label the eigenvectors from the lowest to the highest energy. The Berry connection and Berry curvature for each band can be defined as
$A_{\mu}^m(\vk)=i\ep{\psi_m|\frac{\p}{\p k_{\mu}}|\psi_m}\text{ and } F^m(\vk)=\frac{\p A_y^m}{\p k_x}-\frac{\p A_x^m}{\p k_y},$
with $\mu=x,y$. Furthermore, the Chern number for the lower two occupied bands is given by
\be
C=\frac{1}{2\pi}\sum_{m=1,2}\int d^2k F^m(\vk),
\ee
where the integration is over the whole infinite 2D k-space. In the numerical calculations, a fairly large momentum cutoff is good enough to guarantee that the resulting Chern number is quantized as expected. For the model of Eq.(\ref{Hk}), it is found that the Chern number its valley dependent, which reads
\be
C_{K\xi}=\xi \mbox{sign}(V)
\ee
where $\mbox{sign}(x)=\pm1$ for a positive/negative $x$, and it means that $C_K=\pm1$ with the sign depending on the sign of applied bias. Since this Chern number $C_K$ its valley dependent, it is also called the valley Chern number~\cite{ZhangF13}.

\begin{figure}
\centerline{\includegraphics[width=1.0 \columnwidth]{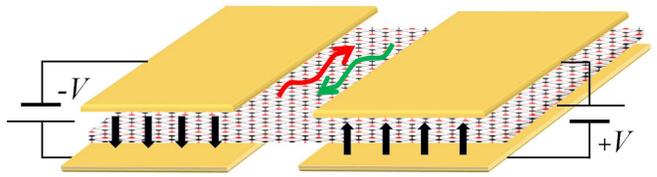}}
\caption{Schematic representation of gated BLG for the creation of a kink potential. The vertical electric field generated by the gated voltage applied to the upper and lower layers with opposite sign, is indicated by the black arrows. Red and green wavy arrows represent counterpropagating chiral modes localized to the domain wall.}
\label{fig-schematics}
\end{figure}

For the DW region formed by potential, the electric field usually depend on spatial position, $V(x)$. Fig. ~\ref{fig-schematics} illustrates schematically the topological chiral modes in BLG. In this paper, we mainly focus on a pair of kink and anti-kink DWs in the BLG for one valley K, as a result of requirement of the periodic system. For a fixed layer, the electric potential is initially positive and switch to negative at the kink DW. Then it changes back to positive at the anti-kink DW. This potential can be written as $V(x)=V_0[ \tanh(\frac{x-x_1}{w})- \tanh(\frac{x-x_2}{w})-1 ]$. Here $V_0$ is the overall magnitude of bias voltage and $w$ denotes the width of region in which the potential switches its sign in each layer. Since the translational symmetry is preserved along the $y$ direction, we can keep the momentum $k_y$. But we have to represent $k_x$ as a differential operator $-i\p_x$ in real space along the $x$ direction. For concreteness, we assume that the system is located inside a finite interval $-L/2<x<L/2$ with periodic boundary conditions as $\psi(-L/2)=\psi(L/2)$. Because $-i\p_x$ will change sign under complex conjugation, we find that particle-hole symmetry is broken in this real space model.

The above 4-band continuum model has particle-hole symmetry. In this form, it is easy to see that
\be
U^{\dagger}H^*U=-H,
\ee
with $U=\tau_1\otimes\sigma_2$. This means that if $\ket{u_n}$ satisfies $H\ket{u_n}=E_n\ket{u_n}$, one also has $H(U\ket{u_n}^*)=-E_n(U\ket{u_n}^*)$. Therefore the spectra are symmetric about $E=0$.

In numerical calculation, one can discretize the $x$ interval with $N$ lattice sites. Then
the 4-band model can be written as
\be
&&H=-\frac 12 \tau_3\otimes \sigma_0\otimes V_m-iv_F\tau_0\otimes\sigma_1\otimes h_1\nonumber\\
&&\qquad +v_F k_y \tau_0\otimes\sigma_2\otimes h_0\nonumber\\
&&\qquad +\frac{t_{\bot}}2(\tau_1\otimes\sigma_1+\tau_2\otimes\sigma_2)\otimes h_0
\label{Hr}
\ee
Here $V_m=V(x_i)\delta_{ij}$, $h_0=\delta_{i,j}$ and $h_1=\frac{1}{2\Delta x}(\delta_{i,j-1}-\delta_{i,j+1})$ with $i,j=1,\cdots,N$ and $\Delta x=L/N$. Note that in Eq.(\ref{Hr}) the derivative is approximated by a symmetric difference operator as $\p_x\to h_1$, thus we can call it symmetric difference model. As discussed before, the differential operator breaks the particle-hole symmetry and one might expect that the spectrum is asymmetric about $E=0$. But the use of symmetric difference actually introduce an extra symmetry to the system and make its spectrum symmetric again. To be more explicit, one can introduce a spatial symmetric transformation as $U_1=(-1)^{i+1}\delta_{i,N-j+1}$. It is easy to verify the following properties as $U_1^{\dagger}V_mU_1=-V_m$ and $U_1^{\dagger}h_1U_1=h_1$. With this matrix $U_1$, one can show the Hamiltonian of Eq.(\ref{Hr}) has a generalized particle-hole symmetry as
\be
U_2^{\dagger}H^*U_2=-H,
\ee
with $U_2=\tau_3\otimes\sigma_0\otimes U_1$. Therefore, the Hamiltonian of Eq.(\ref{Hr}) also has a spectrum symmetric about $E=0$. The numerical results suggest that there appears several chiral gapless modes localized at the DW. Due to the above generalized particle-hole symmetry, there will be two branches of chiral modes propagating along the two opposite directions. The reason for the appearing of two branch of fermions with opposite chirality is the so-called fermion doubling problem first encountered in lattice Quantum Chromodynamics~\cite{Kogut,Creutz}.

To avoid this fermion doubling problem~\cite{Tworzyd08,Alexis12,Messias12}, we follow the suggesting of Stacey~\cite{Stacey82} to replace the symmetric difference $h_1$ by the the forward difference $h_2=\frac{1}{\Delta x}(\delta_{i,j-1}-\delta_{i,j})$. Consequently, the 4-band BLG model with a lattice realization along $x$-direction can be written as
\be
&&H=-\frac 12 \tau_3\otimes \sigma_0\otimes V_m-iv_F\tau_0\otimes\sigma^+\otimes h_2\nonumber\\
&&+iv_F\tau_0\otimes\sigma^-\otimes h_2^T+v_F k_y \tau_0\otimes\sigma_3\otimes h_0\nonumber\\
&&+\frac{t_{\bot}}2(\tau_1\otimes\sigma_1+\tau_2\otimes\sigma_3)\otimes h_0,
\label{Hf}
\ee
where we defined $\sigma^{\pm}=(\sigma_1 \pm i\sigma_2)/2$ for convenience. Comparing to the Eq.(\ref{Hr}), we will see in the next section that there is indeed no fermion doubling problem for the model of Eq.(\ref{Hf}). We will call Eq.(\ref{Hf}) as the forward difference model since one of the derivative is approximated by the forward difference operators.

In experiments, one is more interested in directly measurable quantities as optical conductivity. The non-local optical conductivity can be computed from the Kubo formula~\cite{jiang17,LuoY20} as
\be
&&\sigma_{ab}(x_1,x_2)=\nonumber\\
&&\frac{i g_s}{4 \pi^2}\int d^2\mathbf{k}
\Bigg(\sum_{m\neq n}\frac{f(E_m)-f(E_n)}{(E_m-E_n)(\omega+i\eta-(E_m-E_n))}\nonumber\\
&&+\sum_{m} \frac{d f(E_m)}{d E_m}\frac{1}{\omega+i\eta}\Bigg)\ep{u_m(x_1,k_y)|v_a|u_n(x_1,k_y)}\nonumber\\
&&\qquad\times\ep{u_n(x_2,k_y)|v_b|u_m(x_2,k_y)},
\label{Eq-sig-nonlocal}
\ee
where $g_s=2$ denotes the spin degeneracy in graphene, $f(E_m)$ is the Fermi-Dirac distribution $f(E_m)=1/(1+e^{(E_m-\mu)/(k_B T)})$. The velocity operator can be obtained from
$v_a=\dfrac{\p H}{\p k_a}$ where the spatial indices $a,b=x,y$. For the model of Eq.(\ref{Hr}), we have $v_x=v_F\tau_0\otimes\sigma_1$.
The local optical conductivity can be obtained as
\be
\sigma_{ab}(x)=\int dx'\sigma_{ab}(x,x')
\label{Eq-sig-local}
\ee

\section{Results and discussions}

\subsection{Band structure}

As discussed in the previous section, the valley Chern numbers of the BLG are $\pm1$ with the sign determined by the sign of the applied bias potentials. If the applied bias potential forms a DW like configuration, then the valley Chern number will jump from $+1$ to $-1$ across DW. According to the bulk-edge correspondence, there will appear two chiral modes localized around the DW. This is confirmed by numerical study of the band structure of the 4-band continuum model with a DW like bias potential.

\begin{figure*}%[H]
\centering
{\includegraphics[width=0.49\textwidth]{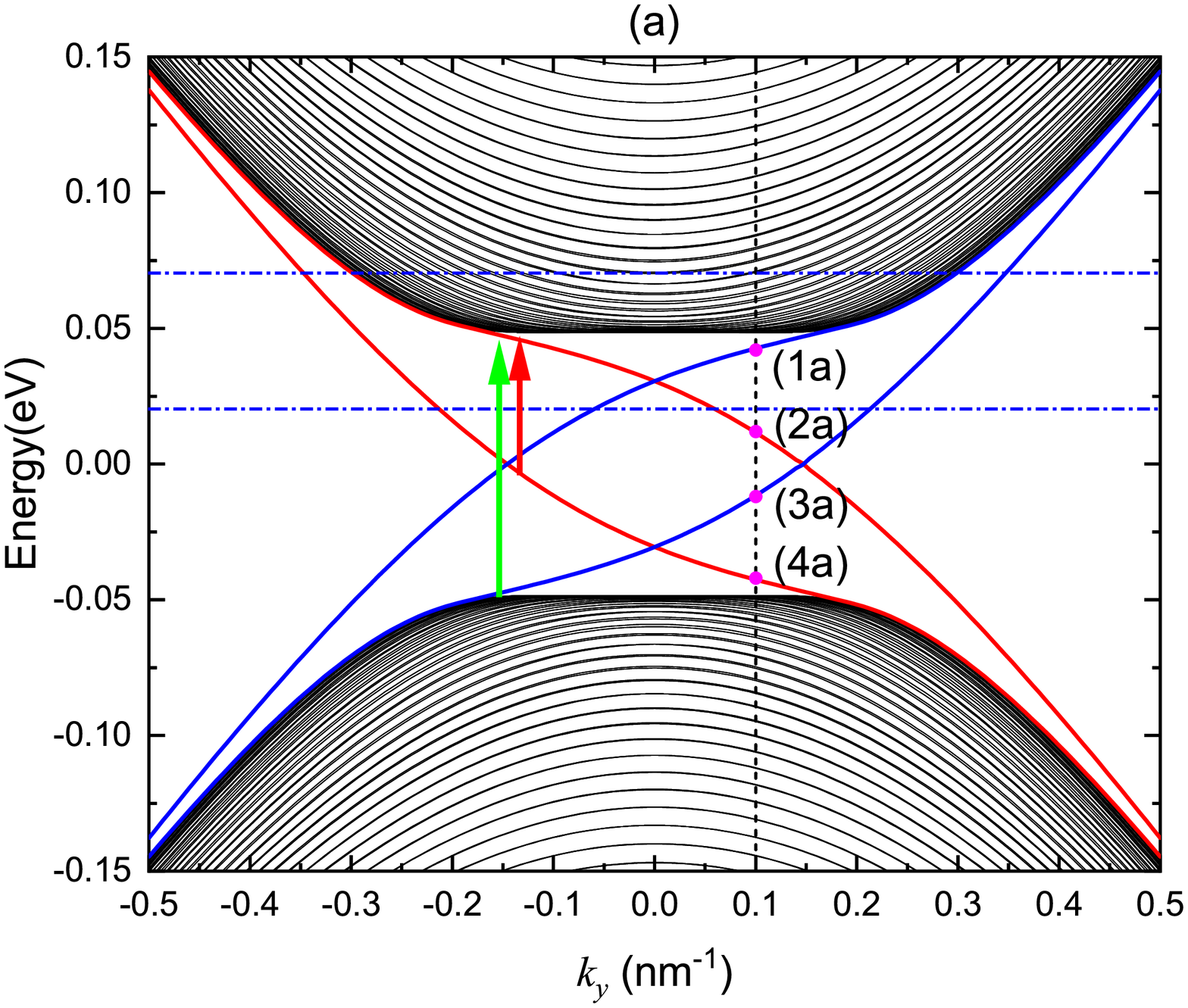}}
\centering
{\includegraphics[width=0.49\textwidth]{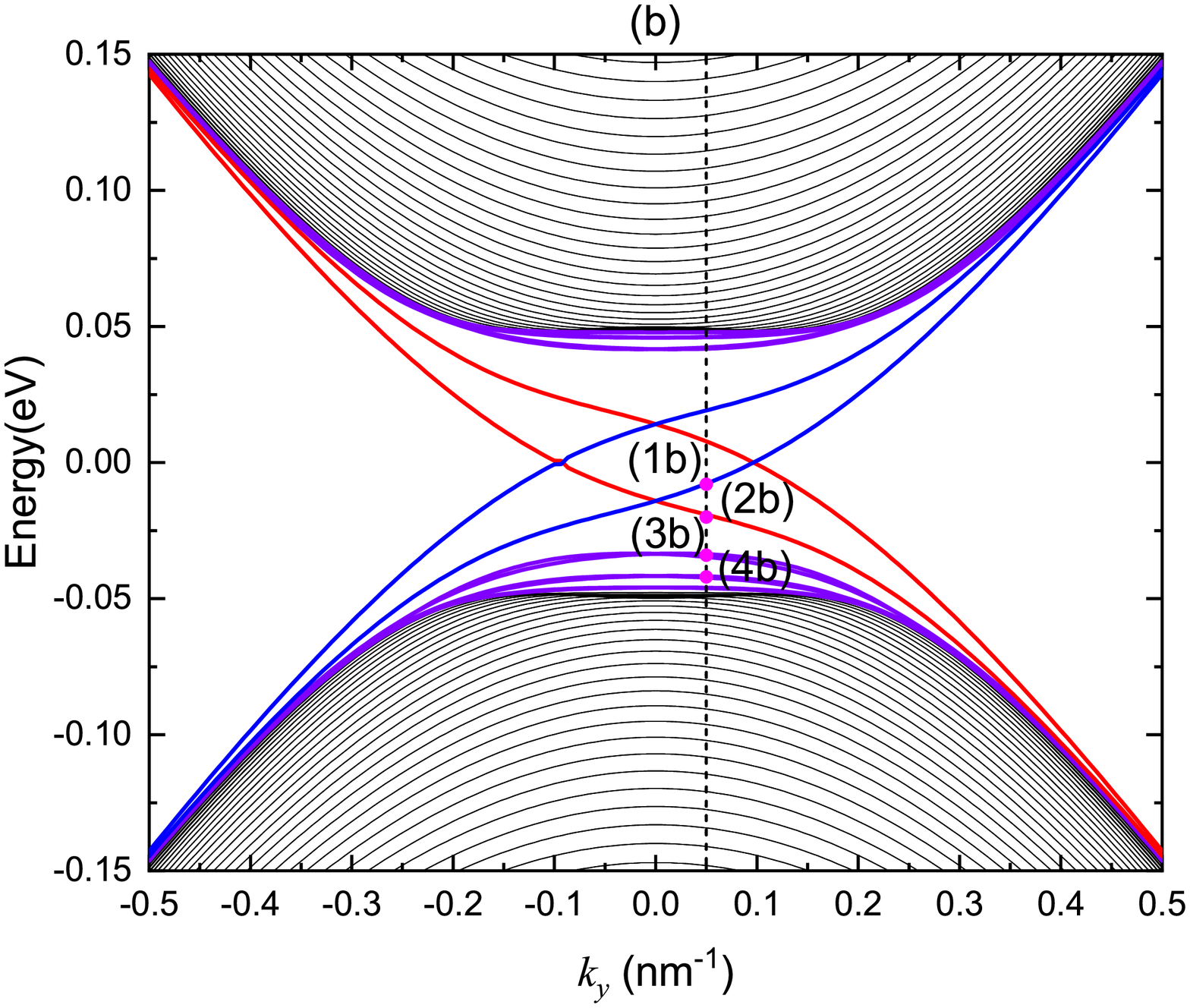}}
\centering
{\includegraphics[width=0.49\textwidth]{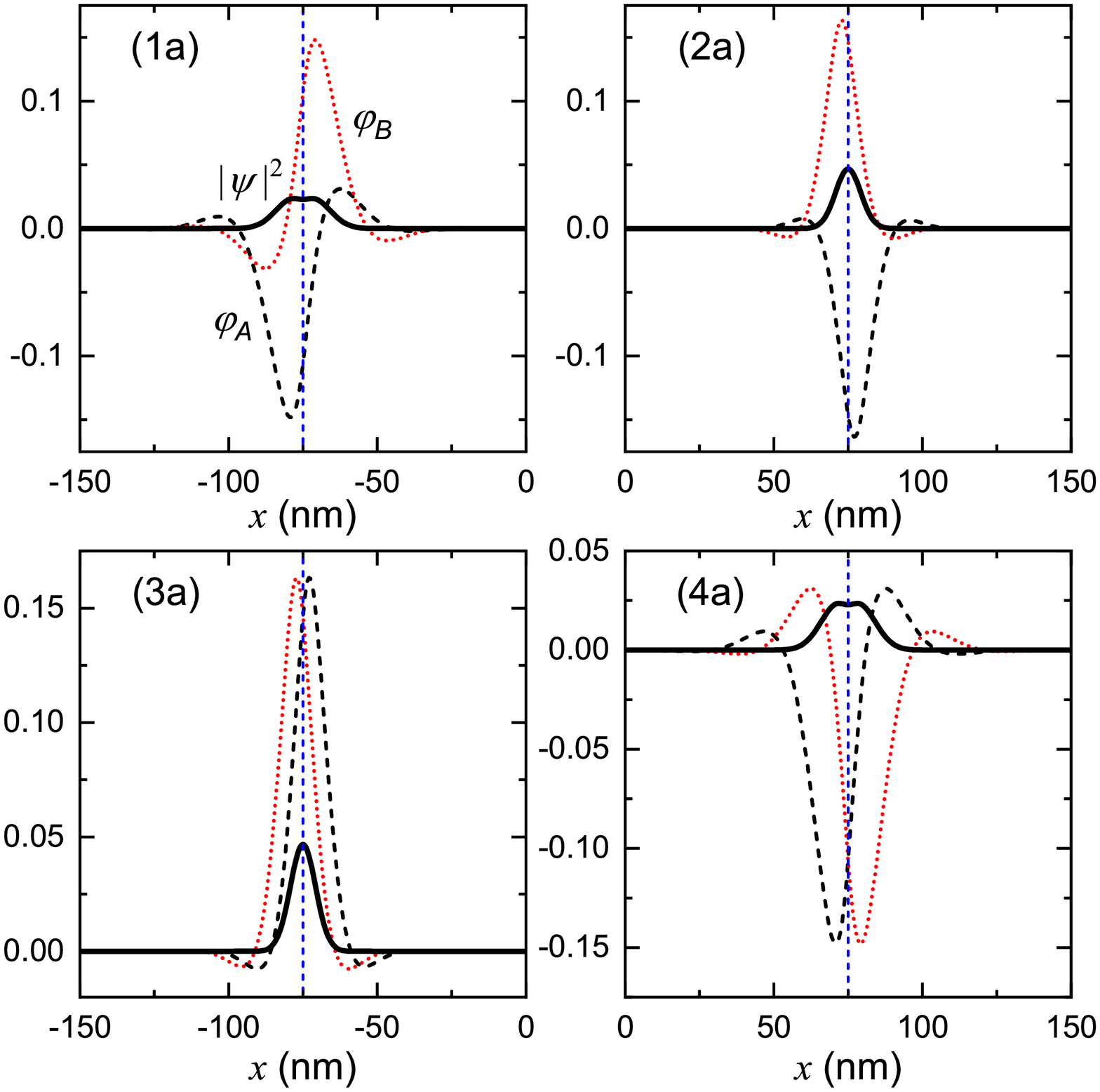}}
\centering
{\includegraphics[width=0.49\textwidth]{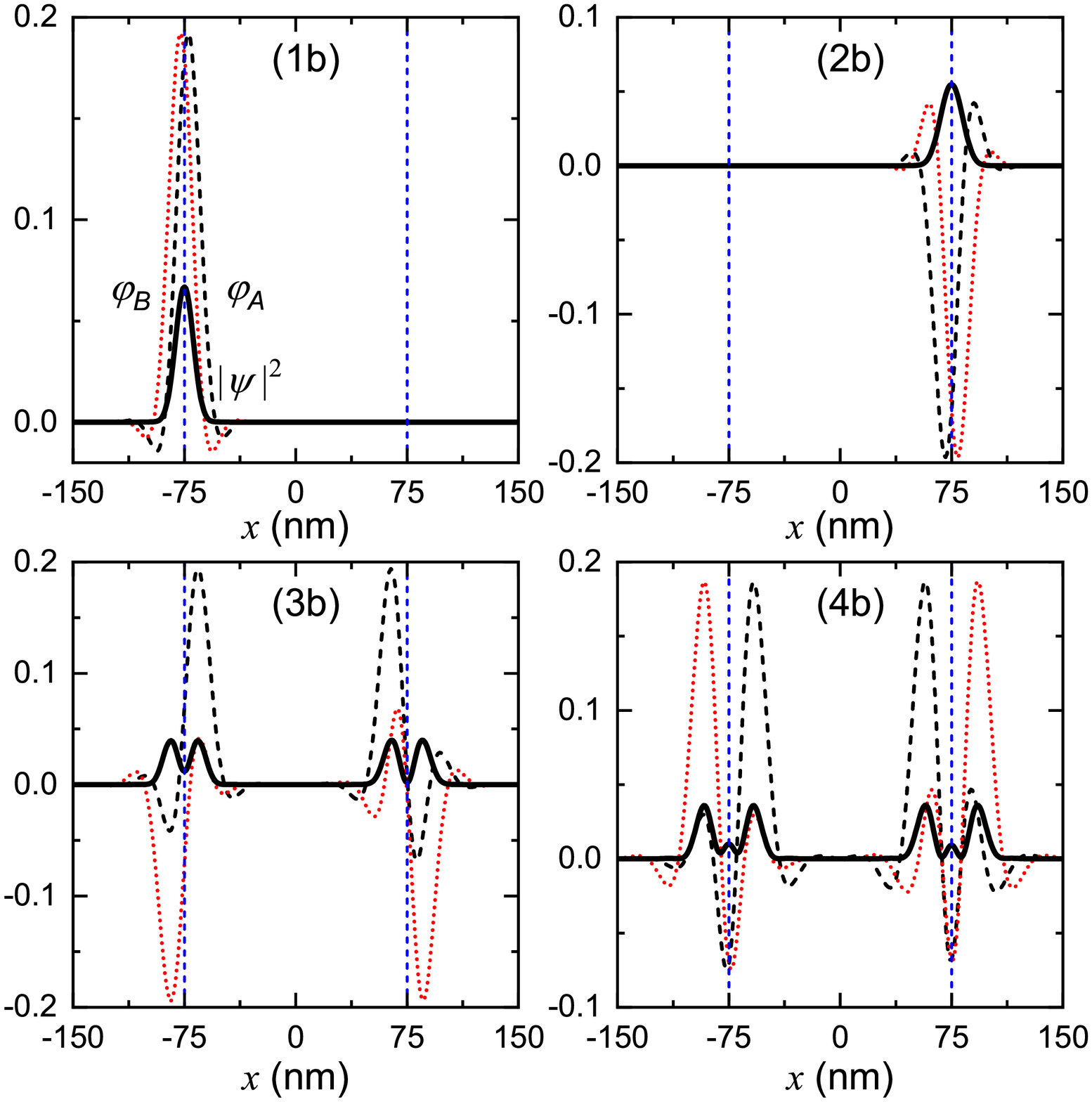}}
\caption{Energy spectrum, wave functions and probability densities of a kink-antikink profile with $V_0=0.1$ V for one valley. (a): Energy spectrum of BLG with a kink-antikink potential for $V_0=0.1$ V, $w=2$ nm. The black dashed line indicates $k_y=0.1\text{ nm}^{-1}$. (b): Energy spectrum of BLG with a kink-antikink potential for $w=15$ nm. The black dashed line indicates $k_y=0.05\text{ nm}^{-1}$. (1a)-(4a): The real parts of the wave functions and probability densities of the states corresponding to the four pink points in panel (a). (1b)-(4b): The real parts of the wave functions and probability densities of the states corresponding to the four pink points in panel (b). The blue dashed lines indicate the center of the two DWs.}
\label{fig-wavef}
\end{figure*}

In Fig.~\ref{fig-wavef}(a), the energy spectrum of Eq.(\ref{Hf}) is plotted as a function of $k_y$. The parameters used in this calculation is displayed in the figure caption. Note that there are two pairs of mid-gap states (the red and blue solid lines) connecting the valence and conduction band. Since the two pairs of gapless modes propagate in the opposite direction, it seems that there may be a fermion doubling problem. To investigate the physical origin of chiral states, we plot the real part of the wave functions of sublattice A (black dashed curves) and B (red dotted curves) and the probability density for the states indicated by the pink points in Fig.~\ref{fig-wavef}(a), corresponding to $k_y=0.1 \text{ nm}^{-1}$. The wave functions are the eigenvectors obtained by diagonalization of the Hamiltonian Eq.(\ref{Hf}). According to the previous results~\cite{Martin08,Zarenia11}, we notice that the wave function have prominently A1 and B2 character at low energies and constant gate voltage. Obviously, the electron states are localized at the left DW region with $x=x_1=-75$ nm, as shown in Fig.~\ref{fig-wavef}(1a) and (3a). In other word, the chiral states at the left DW region are represented by the blue curves in panel (a). Meanwhile, the red curves also indicate the localized states stay at the right DW region with $x=x_2=75$ nm, in Fig.~\ref{fig-wavef}(2a) and (4a). Therefore, these gapless modes actually are localized at the different DW regions with $x=x_1$ and $x_2$ and there is indeed no fermion doubling problem, which corresponds to the two blue and red curves, respectively. Since the two DW regions have opposite orientations, the two pairs of gapless modes should propagate in the opposite directions. In addition, the propagation direction of the gapless states can be reversed by reversing the sign of electric field $V(x)$, are indicated by the red and blue curves in plane (a).

We notice that the peak shape of probability density, $|\Psi|^2=|\Psi_A|^2+|\Psi_B|^2$, in panel (1a) and (4a) is slightly different from that of panel (2a) and (3a). The peaks of (1a) and (4a) are wider and almost split into double peaks. This is because that the energy of the states of (1a) and (4a) are much more closer to the continuum spectrum than that of (2a) and (3a). As the topological states approach to the bulk states, the confining potential is weakened and the wave function is less localized than before.

Next, we investigate the energy spectrum for a broader kink potential profiles, i.e. $w=15$ nm. Fig.~\ref{fig-wavef}(b) shows the band structure of gated BLG with $V_0=0.1$ V, in addition to the topological states, several branches of states split off from the bulk, which is similar to the results reported in \cite{Zarenia11,jiang17,LiJ16}. Here we will refer to these states as bound states. In order to understand the physical origin of these bound states, we also plot the wave functions of the two oppositely propagating topological states and the bound states of electronic band for $k_y=0.05\text{ nm}^{-1}$ in Fig.~\ref{fig-wavef}(1b)-(4b). The bound states are also located around $x=x_1$ or $x_2$, but these states are more extended and their amplitudes cross zero near the center of the DW region. In panel (3b) and (4b), we find that these bound states are localized at both kink and anti-kink DW regions since these two states are very close to each other. While the chiral states are only confined at one DW region as shown in (1b) and (2b). Meanwhile, the probability distribution of bound states is quite different from that of topological states since these two types of states are quite different topological features. As pointed out in \cite{Zarenia11}, the bound states are no longer unidirectional and non-chiral. They have a quasi-1D free-electron type of spectrum and are less localized at the DW region as compared to the chiral states.  Their probability distribution appears similar to those of excited states.

\subsection{Optical conductivity}

\begin{figure}%[H]
\centering
{\includegraphics[width=\columnwidth]{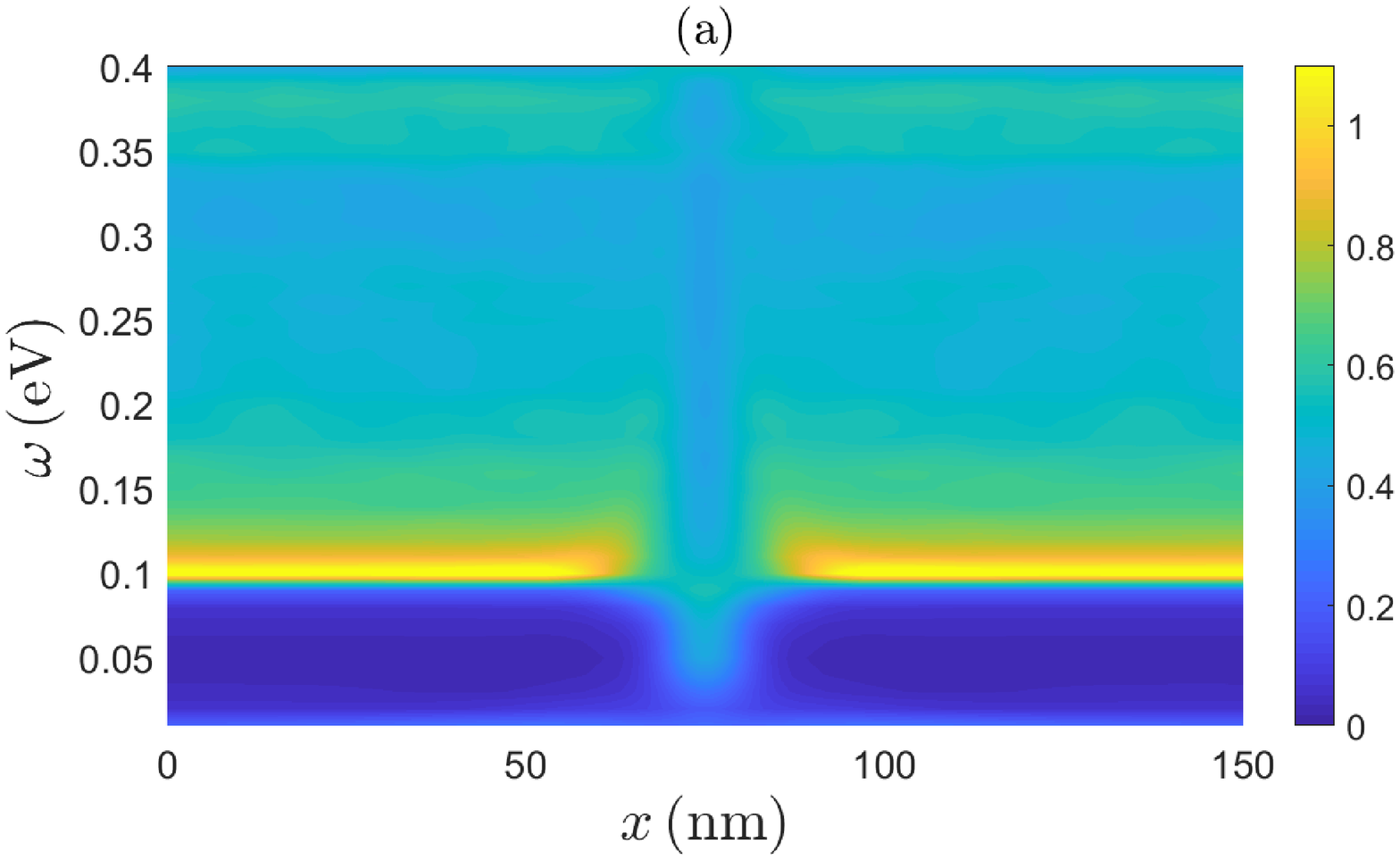}}
\centering
{\includegraphics[width=\columnwidth]{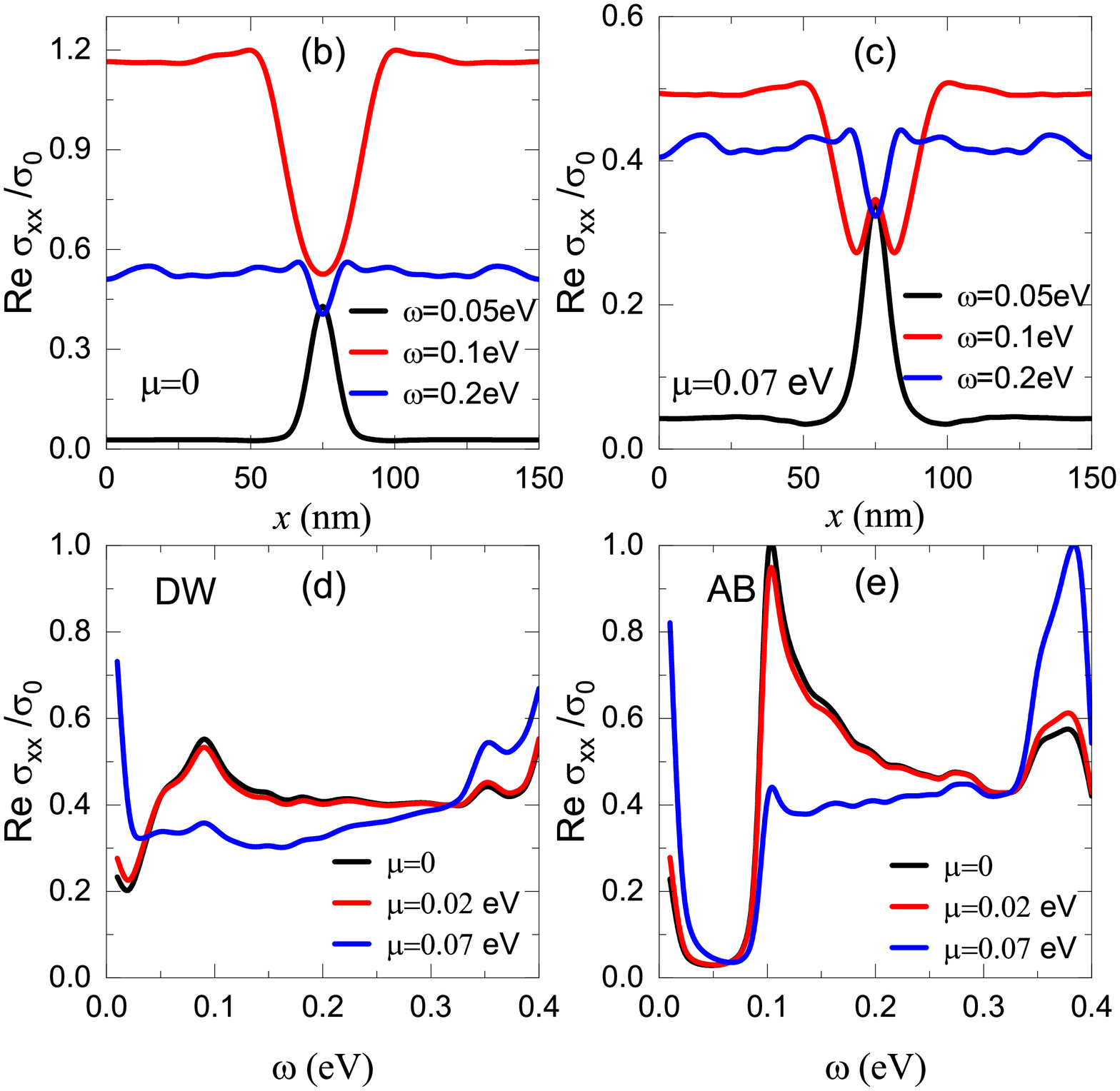}}
\caption{The spatial and frequency dependence of the local optical conductivity of BLG with DW width $w=2$ nm. (a) The real part of the local optical conductivity in unit of $\sigma_0=e^2/\hbar$ as a function of spatial coordinate and frequency. The intensity of optical conductivity is indicated by the color. (b) The spatial distribution of the real part of the optical conductivity for $\omega=0.05, 0.1, 0.2$ eV, at $\mu=0$. (c) The same as (b) but with $\mu=0.07$ eV. (d) The real part of the optical conductivity at the center of DW region as a function of frequency for the different chemical potentials $\mu=0, 0.02, 0.07$ eV. (e) The same as (d) but for the Bernal stacking region of BLG. All quantities were calculated at $V_0=0.1$ V, $t_{\bot}=0.39$ eV, $T=300$ K, and the damping rate is taken to be $\eta=3$ meV.}
\label{fig-oc-L2}
\end{figure}

We now present the results of the local longitudinal optical conductivity $\sigma_{xx}$ of BLG with a DW like bias, which is computed by the Kubo formula of Eq.(\ref{Eq-sig-nonlocal}) and (\ref{Eq-sig-local}). Fig. \ref{fig-oc-L2}(a) shows the frequency dependence of the real part of the local optical conductivity Re $\sigma_{xx}$ in gated BLG with the parameters $w=2$ nm, $V_0=0.1$ V, $\mu=0$. The optical conductivity are scaled by $\sigma_0 = e^2/\hbar$, which is four times the background optical conductivity of monolayer graphene. This includes the spin degeneracy factor $g_s=2$ in graphene since the charged impurity scattering does not mix spins. We notice that the longitude optical conductivity of the DW region is quite different from of that AB region. A strong peak in the optical conductivity away from DW region is visible near $\omega=0.1$ eV. And in the DW region, a peek occurs at the energies below 0.1 eV and a dip appears at $\omega >0.1$ eV.

In order to illustrate the above features more quantitatively, in Fig.~\ref{fig-oc-L2}(b)-(e), we plot Re $\sigma_{xx}$ of gated BLG in the DW and AB region as a function of frequency and spatial coordinate at a series of fixed chemical potential. Fig.~\ref{fig-oc-L2}(b) shows the spatial dependence of the local optical conductivity for several incident photon frequencies at charge neutral point. As illustrated in panel (a), there is a prominent peak located at the DW region at $\omega<0.1$ eV due to the optical transitions between the chiral states and the conduction bands, which enhances the local optical conductivity. As energy increases and exceeds band gap of system, the longitude optical conductivity near the DW region drops sharply which gives rise to a valley-like profile. When the energy is exactly equal to the size of band gap, the contrast between the local optical conductivity at the center of the DW region and that at the AB region is the most evident. In other word, the size of band gap of gated BLG can be easily estimated by the profile of local optical conductivity in real space. As $\omega$ increases, the contrast of the optical conductivity between the DW and AB region become much smaller, as indicated by the blue curve in panel (b). At $\omega=0.2$ eV, the optical conductivity at the AB region is roughly equal to $0.5\sigma_0$, which is twice the optical conductivity of monolayer graphene. For a non-zero chemical potential $\mu=0.07$ eV, where the bulk states are occupied, the dependence of local optical conductivity mentioned above is similar as shown in Fig.~\ref{fig-oc-L2}(c).

Interestingly, the value of optical conductivity at center of DW region varies very little. In order to obtain more information, we plot the optical conductivity at DW region as a function of incident photon frequency for several typical chemical potentials in panel (d). At low energy, there are a shoulder and peak at $\omega=0.05, 0.09$ eV, respectively. We find that the shoulder at $\omega=0.05$ eV corresponds to the contribution of transition between the two topological states as marked by the red arrow in Fig.~\ref{fig-wavef}(a). And the optical conductivity peak at $\omega=0.09$ eV is associated with the optical transition between upper topological states and continuum band as marked by the green arrow in Fig.~\ref{fig-wavef}. These two absorption peaks appear because the chiral state near the continuum modes become less localized at the DW region, which are already illustrated by the wave functions in Fig.~\ref{fig-wavef}(1a) and (4a). When $\mu$ is increasing but still inside the band gap, the optical conductivity curve will stay almost the same. On the other hand, when $\mu$ is outside the band gap, these two peaks of optical conductivity of the DW region disappear. Therefore, the frequency dependence of the optical conductivity can give us more evidence of localized states at DW region.

Fig.~\ref{fig-oc-L2}(e) shows the optical conductivity of BLG at the uniform AB region as a function of frequency for the different chemical potentials. We will call Re~$\sigma_{xx}$ versus $\omega$ curve as the conductivity spectrum. The absorption peak appears at $\omega=0.1$ eV due to the size of band gap $\Delta=0.1$ eV, which can be used as an evidence to detecting the band gap. Another prominent absorption peak appear at $\omega=0.39$ eV, which is consistent with the inter-layer coupling constant $t_\perp=0.39$ eV. When chemical potential enters into bulk bands, for example $\mu=0.07$ eV, the absorption peak of optical transition associated with the band gap is significantly decreased, but the band gap can still be clearly determined. Moreover, the intensity of optical conductivity associated with the inter-layer coupling is significantly increased, with a peak value of roughly $\sigma_0$. Therefore, we can attribute the huge difference between the conductivity spectrum of the DW and AB regions to the different topological properties of eigenmodes.

\begin{figure}%[H]
\centering
{\includegraphics[width=\columnwidth]{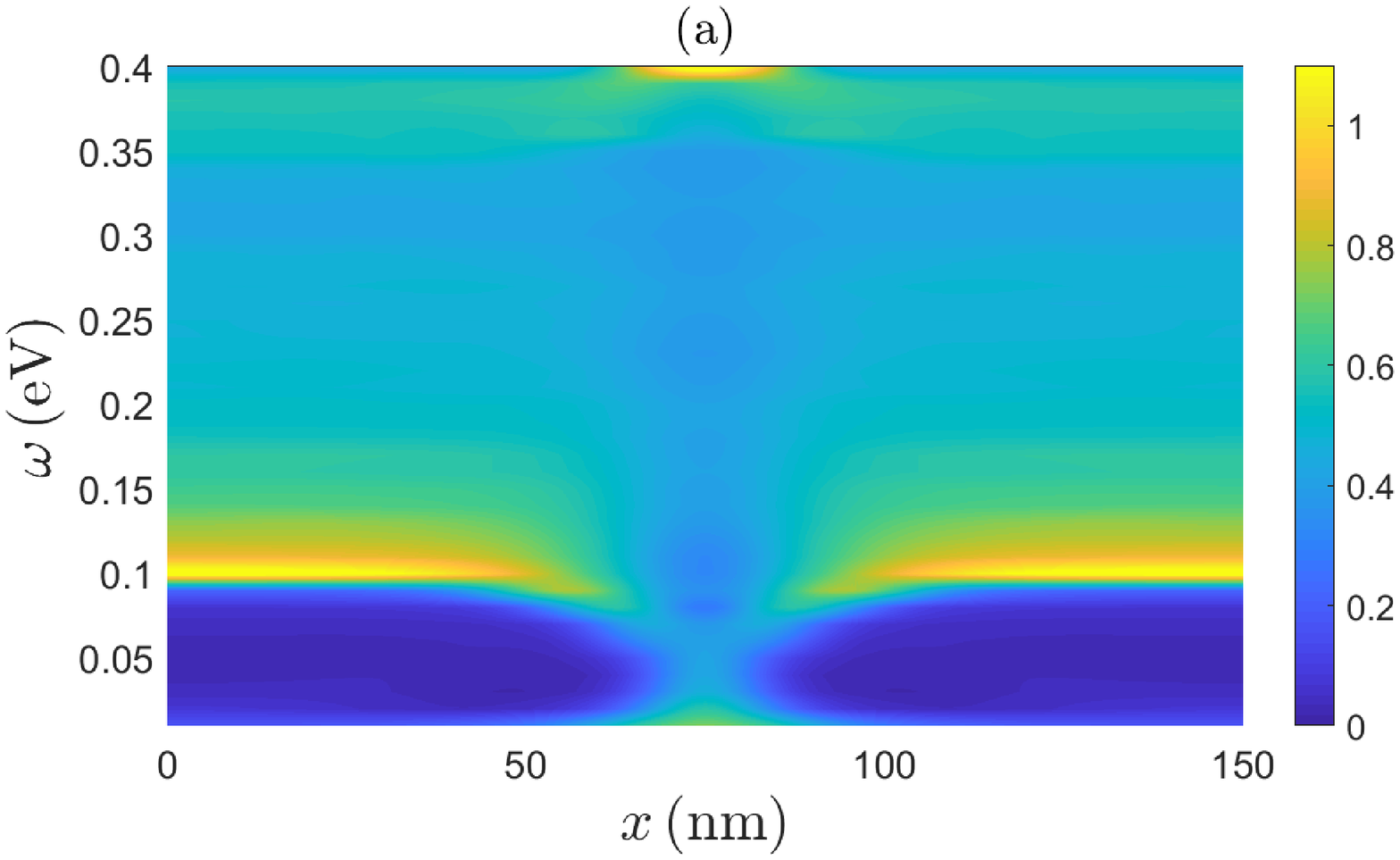}}
\centering
{\includegraphics[width=\columnwidth]{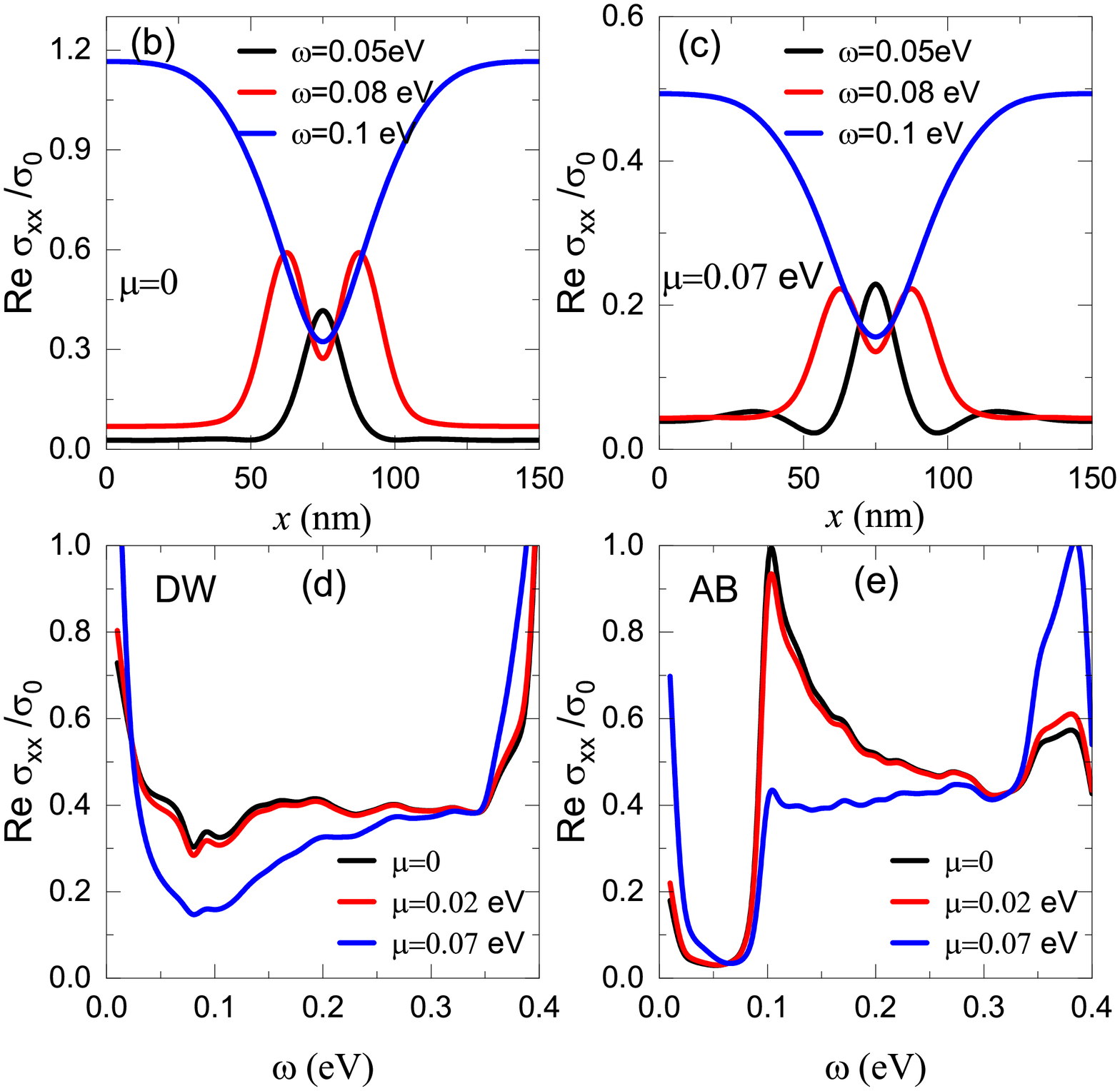}}
\caption{The spatial and frequency dependence of the local optical conductivity of BLG with DW width $w=15$ nm. (a) The real part of the local optical conductivity in unit of $\sigma_0=e^2/\hbar$ as a function of spatial coordinate and frequency. The intensity of optical conductivity is indicated by the color. (b) The spatial distribution of the real part of the optical conductivity for $\omega=0.05, 0.08, 0.1$ eV, at $\mu=0$. (c) The same as (b) but with $\mu=0.07$ eV. (d) The real part of the optical conductivity at the center of DW region as a function of frequency for the different chemical potentials $\mu=0, 0.02, 0.07$ eV. (e) The same as (d) but for the Bernal stacking region of BLG. All quantities were calculated at $V_0=0.1$ V, $t_{\bot}=0.39$ eV, $T=300$ K, and the damping rate is taken to be $\eta=3$ meV.}
\label{fig-oc-L15}
\end{figure}

Next, we increase the width of the DW region and investigate how the local optical conductivity spectrum changes. As mentioned before, for wide DW region, additional 1D bound states appear. These bound states will affect the spatial distribution of the local optical conductivity because they are less strongly localized at the DW region as compared to the chiral states. To see this difference more clearly, we also plot the frequency and spatial dependence of the optical conductivity of the gated BLG with parameters $w=15$ nm, $V_0=0.1$ V and $\mu=0$ in Fig.~\ref{fig-oc-L15}. In panel (a), we find that there is a transition from a single peak at $\omega=0.05$ eV to twin peaks line shape at frequency around $0.05\text{ eV}<\omega<0.1$ eV  and finally to remarkable valley-like profile at $\omega=0.1$ eV. The single peak is related to the enhanced optical transitions of the chiral states at DW region due to these states are tightly localized to the interface. And for higher frequency, the twin peaks in local optical conductivity is manifest near DW region, which is originated from the transition associated with bound states. This line shape changes are quite distinct from that of Fig.~\ref{fig-oc-L2}. Similar behaviors can also be observed for $\mu=0.07$ eV, as in Fig.~\ref{fig-oc-L15} (c). In other word, we can tell if a bound state is present by the appearance of twin peak line shape from the spatial resolved spectrum of local optical conductivity at one particular frequency.

Furthermore, we also show the local optical conductivity spectrum at DW region (panel (d)) and AB region (panel (e)) for several different chemical potentials. We note that there is a small peek near $\omega=0.09$ eV due to the optical transition between the bound states when the chemical potential stays in the band gap. When $\mu=0.07$ eV, this small absorption peak is suppressed because the transitions among bound states are forbidden according to the Pauli blocking principle. In the AB region, the spectrum of optical conductivity are almost the same as Fig. ~\ref{fig-oc-L2} (e).

\section{Conclusion}

In this work, we have investigated the electronic structure and the spatial distribution of optical conductivity near the DWs formed in the reversal gated BLG. We have demonstrated that the propagating direction of the chiral modes, confined to DW, depend on the sign of electric field and valley index. The origin of the gapless states could be deduced from the amplitude profiles of the wave functions of the gapless modes. In this way, we have determined that the two topological edge modes are located at the DW. Moreover, we find that the conductivity spectrum of the DW and AB regions are distinctly different as a result of the different topological states. When the photon energy below the band gap of system, the local longitudinal optical conductivity of BLG with a reversal bias display a prominent peak located at DW. The enhancement of optical conductivity at DW is owing to the emergence of the topological states. When the energy increases and exceeds the band gap, a valley-like profile of optical conductivity appears at DW region. For a smooth kink potential, twin peaks profile in local optical conductivity revealed that bound states are less strongly localized at DW as compared to the chiral states. Therefore, important features in the band structure such as energy separation between chiral states or bound states, localization of these states and band gap of system can be easily estimated by the profile of local optical conductivity. From our method, we can also describe the spatial distribution of the conductivities for a given valley instead of the combination of the contributions from the two different valleys.

The properties of optical conductivity in real space near DW region can help us to better understand the phenomena associated with topology, such as the propagation of topological DW plasmons~\cite{Song17}. Our results present an analysis of the local optical properties of gapped BLG that not only can provide useful insights for designing low-power topological quantum devices~\cite{Qiao11,Qiao14, Cardoso18}, but also might open up a novel way for exploring of the fascinating edge-state and domain-wall physics in BLG.

\begin{acknowledgments}
We are grateful to Prof. Hao-Ran Chang for valuable discussions. This work was supported by the National Natural Science Foundation of China under Grant No. 11874271 and 11874272. Y.H. is also supported by Science Specialty Program of Sichuan University under Grant No. 2020SCUNL210. We thank the High Performance Computing Center at Sichuan Normal University.
\end{acknowledgments}

%\bibliographystyle{apsrev}
%\bibliography{refer}

\end{document}